\documentclass[aps,prd,twocolumn,groupedaddress,amssymb,eqsecnum,showpacs,epsfig,nofootinbib]{revtex4}
\usepackage{graphicx}
\usepackage{bm}
\usepackage{dcolumn}
\usepackage{amsmath}

\usepackage{amsmath}
\usepackage{amssymb}

\numberwithin{equation}{section}
\usepackage{epsfig}

\def \lleq {\lower0.9ex\hbox{ $\buildrel < \over \sim$} ~}
\def \ggeq {\lower0.9ex\hbox{ $\buildrel > \over \sim$} ~}

\def \omm  {\Omega_{0 {\rm m}}}

\def \beq  {\begin{equation}}
\def \eeq  {\end{equation}}
\def \ber  {\begin{eqnarray}}
\def \eer  {\end{eqnarray}}
\def\bq{\begin{equation}}
\def\nq{\end{equation}}
\def\bqr{\begin{eqnarray}}
\def\nqr{\end{eqnarray}}
\let\l=\left
\let\r=\right

\begin{document}
\newcommand{\newc}{\newcommand}

\newc{\be}{\begin{equation}}
\newc{\ee}{\end{equation}}
\newc{\ba}{\begin{eqnarray}}
\newc{\ea}{\end{eqnarray}}
\newc{\bea}{\begin{eqnarray*}}
\newc{\eea}{\end{eqnarray*}}
\newc{\D}{\partial}
\newc{\ie}{{\it i.e.} }
\newc{\eg}{{\it e.g.} }
\newc{\etc}{{\it etc.} }
\newc{\etal}{{\it et al.}}
\newc{\lcdm }{$\Lambda$CDM }
\newcommand{\nn}{\nonumber}
\newc{\ra}{\rightarrow}
\newc{\lra}{\leftrightarrow}
\newc{\lsim}{\buildrel{<}\over{\sim}}
\newc{\gsim}{\buildrel{>}\over{\sim}}
\title{Topological Quintessence}
\author{Juan C. Bueno Sanchez}\email{jcbueno@fis.ucm.es}
\affiliation{Departamento de F\'isica At\'omica, Molecular y Nuclear, Universidad Complutense de Madrid, 28040 Madrid, Spain}

\author{Leandros Perivolaropoulos}
\email{leandros@uoi.gr}
\affiliation{Department of Physics, University of Ionnina, Greece}

\date{\today}

\begin{abstract}
A global monopole (or other topological defect) formed during a recent phase transition with core size comparable to the present Hubble scale, could induce the observed accelerating expansion of the universe. In such a model, topological considerations trap the scalar field close to a local maximum of its potential in a cosmologically large region of space.  We perform detailed numerical simulations of such an {\it inhomogeneous dark energy} system ({\it topological quintessence}) minimally coupled to gravity, in a flat background of initially homogeneous matter. We find that when the energy density of the field in the monopole core starts dominating the background density, the spacetime in the core starts to accelerate its expansion in accordance to a \lcdm model with an effective inhomogeneous spherical dark energy density parameter $\Omega_\Lambda(r)$. The matter density profile is found to respond to the global monopole profile via an anti-correlation (matter underdensity in the monopole core). Away from the monopole core, the spacetime is effectively Einstein-deSitter ($\Omega_\Lambda(r_{out})\rightarrow 0$) while at the center $\Omega_\Lambda(r\simeq 0)$ is maximum. We fit the numerically obtained expansion rate at the monopole core to the Union2 data and show that the quality of fit is almost identical to that of $\Lambda$CDM. Finally, we discuss potential observational signatures of this class of inhomogeneous dark energy models.
\end{abstract}
%
\maketitle
\section{Introduction}
The simplest cosmological model consistent with the observed accelerating expansion of the universe is the \lcdm model \cite{lcdmrev} which assumes that about $70\%$ of the energy density of the universe consists of a homogeneous and constant in time energy density the {\it cosmological constant}. Despite of its simplicity, this model is faced by theoretical (fine tuning and coincidence problem) and observational challenges (eg the CMB anomalies \cite{Copi:2010na} and large scale velocity flows \cite{Watkins:2008hf,Kashlinsky:2008ut}).

Most models generalizing \lcdm are motivated by the theoretical challenges of the model and they are based on breaking the time translation invariance of the cosmological constant while retaining large scale homogeneity. They involve either dark energy \cite{lcdmrev,quintessence} (a time dependent form of energy with negative pressure) or extensions of general relativity with repulsive gravitational properties on large scales. However, observational challenges of \lcdm (CMB anomalies and large scale velocity flows) appear to hint towards violation of large scale homogeneity\cite{Antoniou:2010gw}. Theoretical models generalizing \lcdm and involving violation of large scale homogeneity include Lemaitre-Tolman-Bondi (LTB) models  \cite{Lemaitre:1933qe} based on large voids of matter and no dark energy \cite{GarciaBellido:2008nz}, anisotropic dark energy \cite{Koivisto:2005mm,Battye:2009ze}, fundamental modifications of large scale topology and geometry \cite{Luminet:2008ew}, a Hubble scale magnetic field \cite{Kim:2009gi}, dark energy with Hubble scale inhomogeneities \cite{Grande:2011hm} etc. A common problem of these models is that most of them lack a well defined causal mechanism that generates the proposed Hubble scale inhomogeneities starting from early time homogeneity. For example the size of the matter void required by LTB models is about $2Gpc$ and is unlikely to produce in the context of any known viable cosmological model.

In this study we extend recent work investigating observational constraints on the existence of a Hubble scale spherically symmetric dark energy inhomogeneity \cite{Grande:2011hm}. Such inhomogeneity was found to be consistent with current observations provided that its radius is larger than about $2Gpc$ \cite{Grande:2011hm}. We point out that such inhomogeneity can be naturally produced by a global monopole with a Hubble scale core formed during a recent phase transition. Such a mechanism, called {\it topological quintessence} has similar features as `topological inflation' \cite{vilenkin1,linde,othtopinf} which may have occurred in the early universe and involves accelerating cosmological expansion triggered in the core of global topological defects when proper conditions are satisfied. The investigation of these conditions and the required values of model parameters in the context of the {\it recent universe} where matter is also present is the goal of the present study.

The structure of this paper is the following: In the next section we set up the basic equations of gravitating global monopoles in an expanding universe with matter and use analytical approximations to derive the parameter values that would lead to accelerating expansion in the monopole core at recent times and on Hubble scales. In section 3 we present a numerical solution of the full system of field equations coupled to gravity and show that indeed accelerating expansion takes place in the monopole core when its energy density starts dominating over matter. We also fit the predicted expansion rate in the core to the Union2 data. Finally in Sec.~\ref{concl} we summarize our main results and discuss future prospects of this project.

\section{Gravitating Global Monopoles and Cosmological Expansion}
Global monopoles \cite{barvil} are formed in the context of scalar field theories when the vacuum manifold $\cal M$ contains surfaces that cannot be continuously shrunk to a point namely when $\pi_2 ({\cal M})\neq I$. The corresponding global field configurations are topologically distinct from the vacuum and have most of their energy localized in a spherically symmetric region of space, the global monopole core. The vacuum energy localized in the global monopole core has negative pressure and therefore can induce accelerating cosmological expansion. In order for this expansion to have interesting observable cosmological consequences two basic conditions must be satisfied:
\begin{itemize}
\item
The scale of the monopole core should be cosmological and comparable to the Hubble scale. This condition is necessary so that the effects of the accelerating expansion are applicable on cosmological scales thus affecting cosmological observations (eg. Hubble diagrams).
\item
The vacuum energy in the core dominates over other forms of energy with non-negative pressure (matter or radiation). This condition is necessary so that cosmological dynamics is determined by the negative pressure of the monopole rather than by the other forms of energy present.
\end{itemize}
These two conditions are easily translated to constraints on the parameters of the field theory that predicts the existence of the global monopole. Consider for example the action
\begin{equation}\label{gmaction}
  S=\int d^4 x \sqrt{-g} \left[\frac{m_{Pl}^{~2}}{16\pi}{\cal R}
     -\frac12(\partial_{\mu}\Phi^a)^2-V(\Phi)  + {\cal L}_m \right]\,,
\end{equation}
where $ {\cal L}_m$ is the Lagrangian density of matter fields,  $\Phi^a ~(a=1,2,3)$ is an $O(3)$ symmetric scalar field and
\begin{equation}\label{gmptl}
V(\Phi)= {1\over 4}\lambda(\Phi^2-\eta^2)^2, ~~
\Phi\equiv\sqrt{\Phi^a\Phi^a}\,.
\end{equation}
The vacuum energy density in the monopole core and the size of the core are determined by the two parameters of the model
$\eta$ (the vacuum expectation value) and $\lambda$ (the coupling constant).  The global monopole field configuration is described by the hedgehog ansatz
\begin{equation}\label{glmonanz}
\Phi^a=\Phi(r,t)\hat r^a\equiv
\Phi(r,t)(\sin\theta\cos\varphi,\sin\theta\sin\varphi,\cos\theta)
\end{equation}
with boundary conditions
\begin{equation}
\Phi(0,t)=0\quad,\quad\Phi(\infty,t) =\eta\,,
\end{equation}
where we have allowed for a time dependence having in mind a cosmological setup of an expanding background.
The general spherically symmetric spacetime around a global monopole may be described by a metric of the form
\begin{equation}\label{ltb}
ds^2=-dt^2+A^2(r,t)dr^2+B^2(r,t)r^2(d\theta^2+\sin^2\theta d\varphi^2).
\end{equation}
In order to derive the cosmological dynamical equations we need the total energy momentum tensor $T_{\mu \nu}$. We assume a cosmological setup at recent cosmological times and therefore $T_{\mu \nu}$ is dominated by the global monopole vacuum energy and matter, ie.
\be
	T_{\mu\nu} = T_{\mu\nu}^{(mon)} + T_{\mu\nu}^{(mat)}\,.
\label{tmntot}
\ee
The energy momentum tensor of the monopole is given by
\begin{equation}\label{ein}
T_{\mu\nu}^{(mon)}=\partial_{\mu}\Phi^a\partial_{\nu}\Phi^a
-g_{\mu\nu}\Bigl[\frac12(\partial_{\sigma}\Phi^a)^2+V(\Phi)\Bigr]\,.
\end{equation}
For example for the monopole energy density we have
\be
\rho^{mon}=T_{00}^{mon}=\l[ {\dot \Phi ^2 \over 2} + {\Phi'^2 \over 2A^2} +{\Phi^2 \over B^2 r^2} + {\lambda \over 4} \l(\Phi^2 - \eta^2 \r)^2 \r]\,.
\label{rhomon}
\ee
The energy-momentum tensor of matter is that of a perfect fluid with zero pressure ($p=0$).
\be
T_{\mu\nu}^{(mat)} = \rho^{mat} u_{\mu} u_{\nu}\,,
\label{tmnmat}
\ee
where $\rho^{mat}=\rho^{mat}(r,t)$ is the matter density and $u^{\mu}$ is the velocity 4-vector of the fluid
\be
u^{\mu} = ({1 \over \sqrt{1-v^2}}, {v \over A \sqrt{1-v^2}}, 0, 0)\,.
\ee
with $v=v(r,t)$ the radial matter fluid velocity.
Extremising  the action (\ref{gmaction}) with the metric (\ref{ltb}), the ansatz (\ref{glmonanz}) and the presence of matter we obtain the dynamical equations
\begin{widetext}
\ba
	- G_0^0 &=&
      	K_2^2 (2K-3K_2^2) - 2{B'' \over A^2B} -{B'^2 \over A^2B^2}
	+2{A'B' \over A^3B} -6{B' \over A^2Br} +2{A' \over A^3r}
	-{1 \over A^2r^2} + {1 \over B^2r^2}  \nonumber\\
	&=& {8\pi \over m_{Pl}^2} \l[ \,{\dot{\Phi}^2 \over 2} + {\Phi'^2 \over 2A^2}
	+{\Phi^2 \over B^2r^2} + {\lambda \over 4}(\Phi^2 - \eta^2)^2
	+{\rho^{mat} \over 1-v^2}\, \r] \,,
\label{dyneq0}
\ea
\ba
	{1 \over 2} G_{01} =
	K_2^{2\prime}  + \l( {B' \over B} + {1 \over r} \r) \l( 3K_2^2 - K \r) =
	{4\pi \over m_{Pl}^2} \l( \dot{\Phi} \Phi' -{v \over 1-v^2}\,A\,\rho^{mat} \r) \,,
\label{dyneq1}
\ea
\ba
	{1 \over 2}\,(G_1^1 + G_2^2 + G_3^3 - G_0^0) &=&
      	\dot{K} - (K_1^1)^2 - 2(K_2^2)^2 \nonumber\\
	&=& {8\pi\over m_{Pl}^2} \l[ \, \dot{\Phi}^2 - {\lambda \over 4}(\Phi^2 - \eta^2)^2
	+ {1 \over 2} {1+v^2 \over 1-v^2}  \rho^{mat}\, \r] \,,
\label{dyneq2}
\ea
and
\be
	\ddot{\Phi} - K\,\dot{\Phi} - {\Phi'' \over A^2}
	- \l( -{A' \over A} + {2B' \over B} + {2 \over r} \r)\,
	{\Phi' \over A^2} + {2\Phi \over B^2 r^2} + \lambda \Phi (\Phi^2-\eta^2) = 0.
\label{dyneqphi}
\ee
\end{widetext}
where prime (dot) denotes differentiation with respect to radius $r$ (time $t$) and
\begin{equation}
K_1^1 = - {\dot{A} \over A}\quad,\quad K_2^2 = K_3^3 = - {\dot{B} \over B}\quad,\quad K = K_i^i\,.
\label{kdef}
\end{equation}
The conservation of the energy momentum tensor for the matter fluid (\ref{tmnmat}) in the metric (\ref{ltb}) leads to the following equations for $v$ and $\rho^{mat}$
\be
\frac{\dot v}{v}=(v^2-1) \frac{\dot A}{A}-\frac{v'}{A}\,,
\label{vcons}
\ee
\be
\frac{\dot \rho^{mat}}{\rho^{mat}}= \frac{\dot v}{v}-2\frac{\dot B}{B} - \frac{v}{A} \frac{(\rho^{mat})'}{\rho^{mat}} -2\frac{v}{A}\frac{B'}{B}-\frac{2 v}{r \; A}
\label{rhocons}
\ee
Notice that in the homogeneous FRW limit ($B(r,t)=A(r,t)=a(t)$, $\Phi(r,t)=\eta$) we obtain the familiar Friedman equations with $v=0$ and the usual conservation of matter equation.

It is straightforward to show \cite{barvil} that for $\rho^{mat}=0$, the system (\ref{dyneq0})-(\ref{dyneqphi}) has a static solution $\Phi(r,t)=f(r)$ that is stable for a range of the parameters $\lambda$ and $\eta$ of the model. It is straightforward to show using (\ref{dyneqphi}) that in a locally flat space ($A\simeq B \simeq 1$), the core scale (where $\Phi/\eta \lsim 1$) of this static global monopole is
\be \delta \simeq \lambda^{-1/2} \eta^{-1}\,, \label{coresc} \ee while the vacuum energy density in this core region is
\be \rho^{core} \simeq \frac{\lambda \eta^4}{4}\,. \label{rhomoncore} \ee
In addition, in the flat space approximation the asymptotic behavior of the static global monopole solution may be found using Eq.~(\ref{dyneqphi}) to be of the form
\ba
\Phi(r) &\simeq &\eta\; r/\delta, \;\;\; r\ll \delta \label{phinear}  \\
\Phi(r) &\simeq & \eta\; \left(1- {1\over (r/\delta)^2}\right), \;\;\; r \gg \delta \,. \label{phiaway}
\ea
The above described flat space approximation is valid as long as $\eta \ll m_{Pl}$. For $\eta$ comparable to $m_{Pl}$ the effects of gravity must be taken into account and the coupled system (\ref{dyneq0})-(\ref{dyneqphi}) needs to be solved. An approximate solution may be obtained valid far away from the monopole core \cite{chovil} ($r \gg \delta$) where $\Phi(r) \simeq \eta$ and $\rho^{mon} \simeq {\eta^2 \over {B^2 r^2}}$. In the coordinates of the metric (\ref{ltb}), this approximate solution is of the form \cite{chovil}:
\ba
A(r,t)&=& 1 \label{scfmonera1} \\
B(r,t)&=& 1 +\sqrt{8\pi {\eta^2 \over m_{Pl}^2}} {t\over r}
\label{scfmonera2}
\ea
where the contribution of matter has been ignored. In the Appendix we show a generalization of eqs (\ref{scfmonera1}), (\ref{scfmonera2}) when matter is also present.

Deep inside the monopole core ($r\ll \delta$), spacetime is approximately homogeneous and we may set $A(r,t)\simeq B(r,t) \equiv a_{in}(t)$. In this limit, Eq.~(\ref{dyneq0}) becomes
\be
3 H_{in}(t)^2 = {8\pi \over m_{Pl}^2} \l( {\lambda \eta^4 \over 4} + \rho^{mat}_{0,in} \l( {a_{0,in} \over a_{in}}\r)^3 \r)\,,
\label{lcdmmon}
\ee
where $a_{0,in}\equiv a_{in}(t_0)$ is the scale factor at the present time $t_0$, $\rho^{mat}_{0,in}$ is the present density of matter in the monopole core and
\be
H_{in}(t)\equiv \l({{\dot a_{in}} \over a_{in}}\r)\,.
\label{hofft}
\ee
Similarly, far away from the monopole core ($r\gg \delta$), spacetime is also approximately homogeneous and we may set $A(r,t)\simeq B(r,t) \equiv a_{out}(t)$. In this limit, Eq.~(\ref{dyneq0}) becomes
\be
3 H_{out}(t)^2 = {8\pi \over m_{Pl}^2}  \rho^{mat}_{0,out} \l( {a_{0,out} \over a_{out}}\r)^3\,,
\label{matmon}
\ee
corresponding to a flat matter dominated universe.

In the context of {\it topological quintessence} a cosmologically large region of space associated with the core of a recently formed global monopole starts an accelerating expansion in accordance with Eq.~(\ref{lcdmmon}).
For proper values of the parameters of the model $\lambda$ and $\eta$ (so that ${\lambda \eta^4 \over 4} \gsim \rho^{mat}_{0,in}$) the local expansion rate predicted by (\ref{lcdmmon}) is identical to that of $\Lambda$CDM.

Around a time $t_0$ when $\lambda \eta^4 \gg \rho^{mat}_{0}$ the vacuum energy dominates over matter, the scale factor may be approximated as
\be
a(t)\simeq a(t_0) e^{H_0 (t-t_0)}\,,
\label{aoftcore}
\ee
where
\be
H_0\equiv H(t_0)=\sqrt{{2\pi \lambda \eta^4}\over {3 m_{Pl}^2}}\,.
\label{h0core}
\ee
The slow evolution of the scalar field $\Phi$ may also be approximately obtained using (\ref{dyneqphi}). For $r \ll \delta$ and in the slow roll approximation this becomes
\be
3 H(t_0) {\dot \Phi} \simeq \lambda \eta^2 \Phi\,,
\ee
where $t_0$ is the present time. This leads to \cite{vilenkin1}
\be
\Phi(r,t)\simeq \Phi(r,t_0) e^{{{\lambda \eta^2}\over 3 H_0} (t-t_0)}\,.
\label{evolphicore}
\ee
In the limit
\be
{{\lambda \eta^2}\over 3 H_0^2} \ll 1
\label{statcond1}
\ee
the field is practically static
on cosmological timescales ($\Delta t \simeq H_0^{-1}$) and the static monopole approximation of Eqs.~(\ref{coresc})-(\ref{phiaway}) holds. Notice however, that despite of the approximately static profile of the scalar field, the expansion rate inside the monopole core (Eq.~(\ref{lcdmmon})) may differ significantly from the expansion rate far away (Eq.~(\ref{matmon})).

For
\be
{{\lambda \eta^2}\over 3 H_0^2} \gsim 1
\label{dyncond}
\ee
the field profile evolution is significant on cosmological timescales. The corresponding condition on the parameters of the model is easily obtained from Eqs.~(\ref{statcond1}) and (\ref{h0core}) as
\be
\eta \gsim  \sqrt{1\over {2\pi}} m_{Pl} \simeq 0.4 \; m_{Pl}\,.
\label{etacondnm}
\ee
This is the condition for breakdown of the static monopole solution. A similar result (derived in a different manner) was also presented in Ref. \cite{vilenkin1} in the context of {\it topological inflation} (see also \cite{linde}) and confirmed numerically in Ref. \cite{maeda} and with the presence of radiation in Ref. \cite{chovil}. The numerically obtained condition for accelerating expansion of the monopole core (in physical coordinates) was found to be $\eta > 0.33 \;m_{Pl}$ in good agreement with Eq.~(\ref{etacondnm}). Notice that in this case there is no condition imposed on the coupling constant $\lambda$.

In the context of late time accelerating expansion in the presence of matter, the two conditions that need to be satisfied for consistency with observations are translated to conditions on the model parameters as follows:
\begin{enumerate}
\item
The scale of the monopole core should be comparable to the Hubble scale ($\delta = \alpha H_0^{-1}$ where $\alpha$ is a parameter whose range is to be determined observationally). This is expressed by the condition
\be
{8 \; \pi \over {3 m_{Pl}^2}}\l({\lambda \; \eta^4 \over 4} + \omm \; \rho_{0cr} \r) \simeq \frac{\lambda \eta^2}{\alpha^2}\,,
\label{condmat1}
\ee
where $\omm \equiv {\rho^{mat}_{0} \over \rho_{0cr}}$ is the present fractional density of matter over the critical density $\rho_{0cr}\simeq 8 \times 10^{-47} \; h^2 \; GeV^4$ inside the core, with $h$ the Hubble parameter in units of $100 km/(sec\cdot Mpc)$.
\item
The vacuum energy density in the monopole core must be large enough to start dominating over the matter energy density today. This condition implies that
\begin{eqnarray}
{\lambda \; \eta^4 \over 4}&\equiv& \Omega_{TQ} \; \rho_{0cr} = (1-\omm) \; \rho_{0cr} \nonumber\\
&\simeq&4 \; (1-\omm) \; 10^{-123} \; m_{Pl}^4 \; h^2\,,
\label{condmat2}
\end{eqnarray}
where  $\Omega_{TQ}=1-\omm\simeq 0.73$ is the normalized vacuum energy density in the monopole core.
\end{enumerate}
Using Eqs.~(\ref{condmat1}) and (\ref{condmat2})  we find
\ba
{\bar \eta}^2 & = &  \frac{3\times (1-\omm)}{2\pi \alpha^2} \label{tqcond1mat} \\
\lambda & = & \frac{64 \times \pi^2 \alpha^4}{9\; (1-\omm)}10^{-123}\; h^2 \,,
\label{tqcond2mat}
\ea
where  ${\bar \eta}\equiv \frac{\eta}{m_{Pl}}$ is the dimensionless vacuum expectation value of the global monopole field potential. The fine tuning of the coupling constant $\lambda$ is worth noting but it is not worse than the fine tuning corresponding to the standard \lcdm model.

We have therefore used analytical arguments to derive the values of the global monopole model parameters that are required to induce the observed accelerated cosmological expansion in the context of topological quintessence. In what follows we solve the system (\ref{dyneq0}) - (\ref{rhocons}) numerically in order to demonstrate explicitly that the topological quintessence mechanism is observationally viable.

\section{Numerical Solution of Matter - Global Monopole System}
\subsection{Method}
In order to solve the system (\ref{dyneq0}) - (\ref{dyneqphi}) numerically we use the following rescaled dimensionless variables:
\[
{\bar r}  \equiv  \sqrt{\lambda} \eta r\quad,\quad {\bar t}  \equiv  \sqrt{\lambda} \eta t
\quad,\quad{\bar \Phi} \equiv \frac{\Phi}{\eta}
\]
\[
{\bar \rho}^{mat} \equiv \frac{\rho^{mat}}{\lambda \eta^4}\quad,\quad {\bar \rho}^{mon} \equiv \frac{\rho^{mon}}{\lambda \eta^4}
\]
and in what follows we omit the bar for simplicity. It is straightforward to express the system of Eqs.~(\ref{dyneq0})-(\ref{rhocons}) in terms of the above dimensionless quantities. The form of the equations remains almost unchanged with the following minor modifications:
\begin{itemize}
\item
We replace all dimensional parameters with the corresponding barred dimensionless while $\lambda$ and $\eta$ get eliminated by the rescaling and are replaced by 1 in all equations.
\item
The factor $\frac{8\pi}{m_{Pl}^2}$ on the righthand side of Eq.~(\ref{dyneq0}) gets replaced by $8\pi {\bar \eta}^2$, and similarly for Eqs.~(\ref{dyneq1}) and (\ref{dyneq2}).
\end{itemize}
We solve the system of Eqs.~(\ref{dyneq1})-(\ref{rhocons}) (in its rescaled form) and use the constraint in Eq.~(\ref{dyneq0}) as a test of the accuracy of the solution. We find that in all runs this constraint is satisfied at a level better than $1\%$. We start the evolution at a rescaled time $t_0$ well in the matter era assuming an initially homogeneous background of matter $\rho^{mat}(r,t_0)=4.5$, which implies that \be \frac{\rho^{mat}(r,t_0)}{\rho^{core}}=18 \label{t0def} \ee where $\rho^{core}\equiv \rho^{mon}(0,t)=\frac{1}{4}$. The initial field profile corresponds to the static monopole solution $\Phi(r,t_0)=f(r)$. We thus use the following initial conditions chosen so that Eq.~(\ref{dyneq0}) is satisfied (all quantities are rescaled dimensionless):
\begin{eqnarray}
\rho^{mat}(r,t_0)&=&4.5\,, \label{rhoinit} \\
v(r,t_0)&=&0\,, \label{rhoinit} \\
t_0 &=& \sqrt{\frac{1}{6 \rho(t_0,r) \pi {\bar\eta}^2}}\,, \label{t0init} \\
A(r,t_0)&=&B(r,t_0)=1\,, \label{abinit}\\
\Phi(r,t_0)&=& f(r)\,, \label{phiinit} \\
{\dot \Phi}(r,t_0)&=& 0\,, \label{dphiinit} \\
K_2^2(r,t_0)&=& -\frac{\eta}{\sqrt{3}}\sqrt{\rho^{mat}(r,t_0)+\rho^{mon}(r,t_0)}\label{k22init}\,, \\
K(r,t_0)&=& -\frac{3 \eta}{\sqrt{3}}\sqrt{\rho^{mat}(r,t_0)+\rho^{mon}(r,t_0)}\,.\qquad\label{kinit}
\end{eqnarray}

These initial conditions correspond to the physical assumption of the formation of a global monopole during a recent phase transition that takes place in a flat homogeneous matter dominated universe. Initial curvature could be introduced by modifying the initial condition for the scale factors $A(r,t_0)=B(r,t_0)=1$. The corresponding boundary conditions used during the evolution are the following:
\ba
\Phi(0,t) &\simeq & 0 \,,\label{phibound1} \\
\Phi'(0,t) &\simeq & f'(r=0)\,, \label{phibound2} \\
K(0,t)&=& 3\; K_2^2(0,t)\,. \label{kbound}
\ea
The present time $t_p$ where the evolution of the above system stops is determined by demanding that
\be
\frac{\rho^{mat}(0,t_p)}{\rho^{core}}=\frac{\Omega_{0m}}{1-\Omega_{0m}}\simeq 0.37\,,
\label{tpdef}
\ee
where we have set $\Omega_{0m}=0.27$. It is now straightforward to evolve the above cosmological system from the initial time $t_0$ to the present time $t_p$. In the next subsection we present the results of this evolution and the comparison with observations. The key assumption in the interpretation of these results is that the core size of the global monopole constitutes a large part of the present Hubble scale as discussed in the previous section.
\subsection{Results}
The only parameter that needs to be fixed in the evolution of above system is the scale of symmetry breaking $\eta$. We have considered values of $\eta$ in the range $\eta \in \left[7\times 10^{-4},0.6\right]$ with no significant change in the form of the rescaled expansion rate profile at recent cosmological times. The results presented here correspond to $\eta = 0.1$ unless otherwise noted. The value of the coupling constant $\lambda$ is fixed implicitly for each $\eta$ by demanding that the energy density at the monopole core is comparable to the present matter density \be \rho^{mat}(0,t_p) = 0.37 \; \rho^{core} = \frac{0.37}{4} \ee (see Eq.~(\ref{tpdef})).

\begin{figure}[htbp]
\epsfig{file=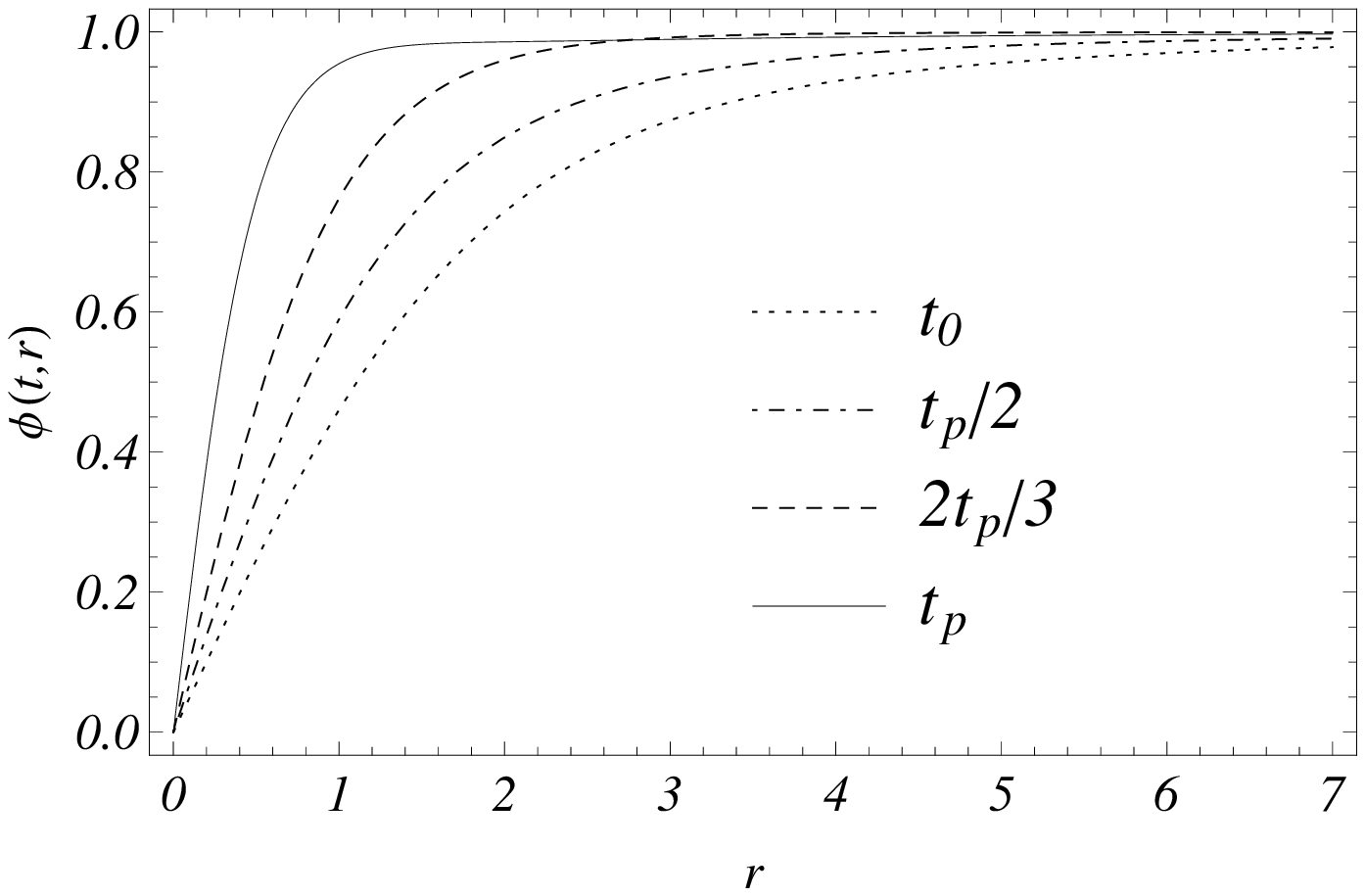,width=8cm}
\epsfig{file=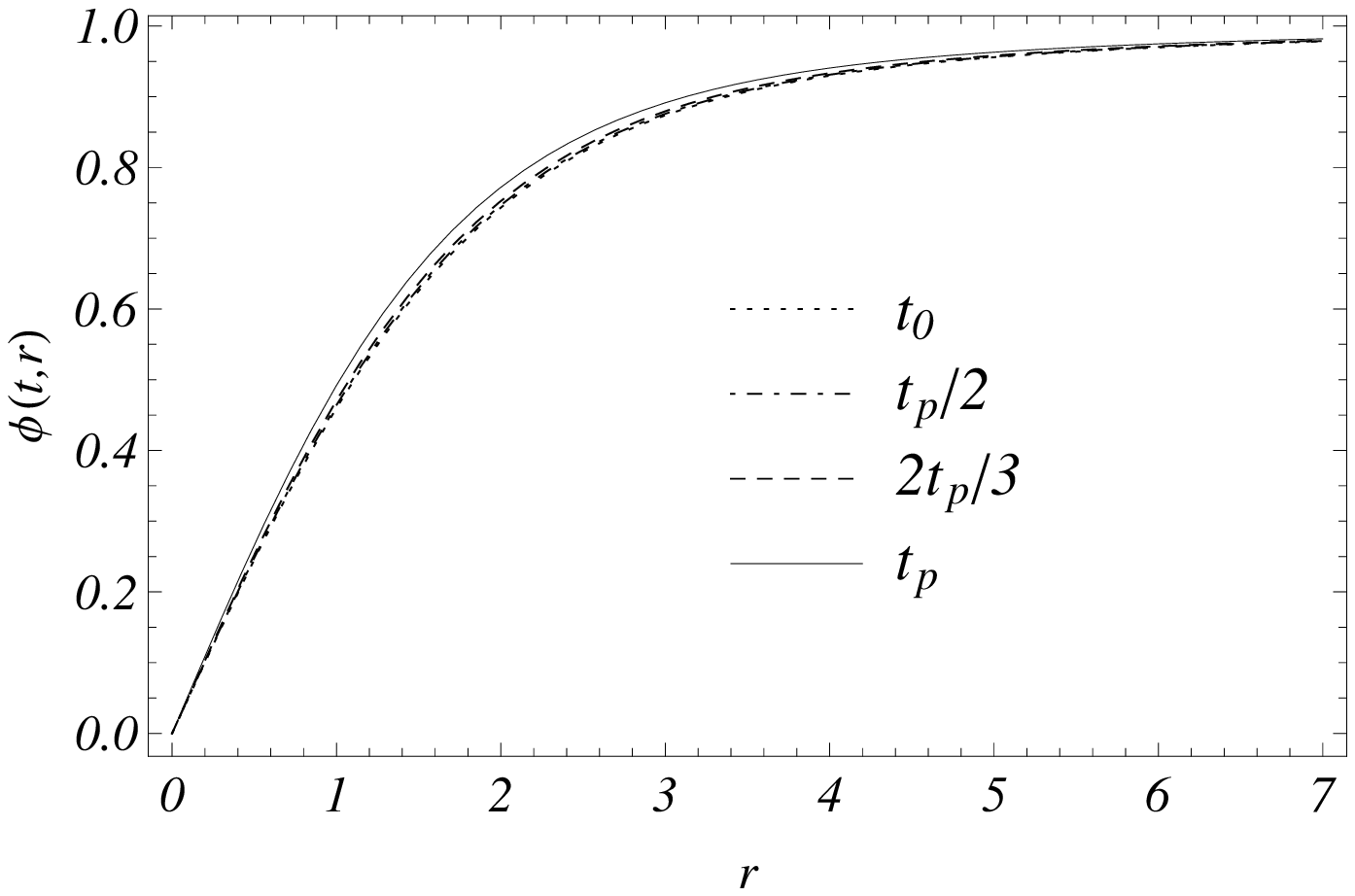,width=8cm}
\caption{Evolution of the scalar field profile for $\eta=0.1$ (top panel) and \mbox{$\eta=0.6$} (bottom panel) from the time $t_0$ up to the present time $t_p$. The profiles shown correspond to $t_0$ (dotted line), $t_p/2$ (dot-dashed line), $2t_p/3$ (dashed line), $t_p$ (solid line) and the profile collapses as time increases. Notice that the evolution gets much slower as we increase $\eta$ because the exponent of Eq. (\ref{evolphicore}) decreases as $\eta^{-2}$ for fixed monopole core density.}\label{fig1}
\end{figure}

In Fig.~\ref{fig1} we show the evolution of the scalar field profile for $\eta=0.1$ and $\eta=0.6$ from the time $t_0$ up to the present time $t_p$. As expected from Eq. (\ref{etacondnm}), for $\eta=0.1$ the field rolls down to its vacuum on a timescale less than $H_0^{-1}$. This roll-down takes place at all comoving radii $r$ and manifests itself as a collapse of the scalar field profile in comoving coordinates. For $\eta=0.6$ the effect is much slower because for constant $\rho^{core}$ (as is the requirement is our case) the exponent of Eq. (\ref{evolphicore}) is written as
\be
\xi(\eta)\equiv{{\lambda \eta^2}\over 3 \; H_0} \Delta t={{4\;\rho^{core}}\over 3 \; H_0} {{\Delta t}\over \eta^2}\sim \eta^{-2}
\label{exponent}
\ee
since in our case $\Delta t \equiv t_p-t_0$, $H_0$ and $\rho^{core}$ are fixed as we vary $\eta$. Thus as we increase $\eta$ by a factor of $6$, the exponent of Eq. (\ref{evolphicore}) decreases by a factor of $36$ and the evolution of $\Phi$ gets suppressed. Indeed we have checked that in our simulations the rescaled exponent of eq. (\ref{evolphicore}) $\frac{t_p-t_0}{H_p}$ varies with $\eta$ as $\eta^{-2}$ as expected from the above arguments.

\begin{figure}[htbp]
\epsfig{file=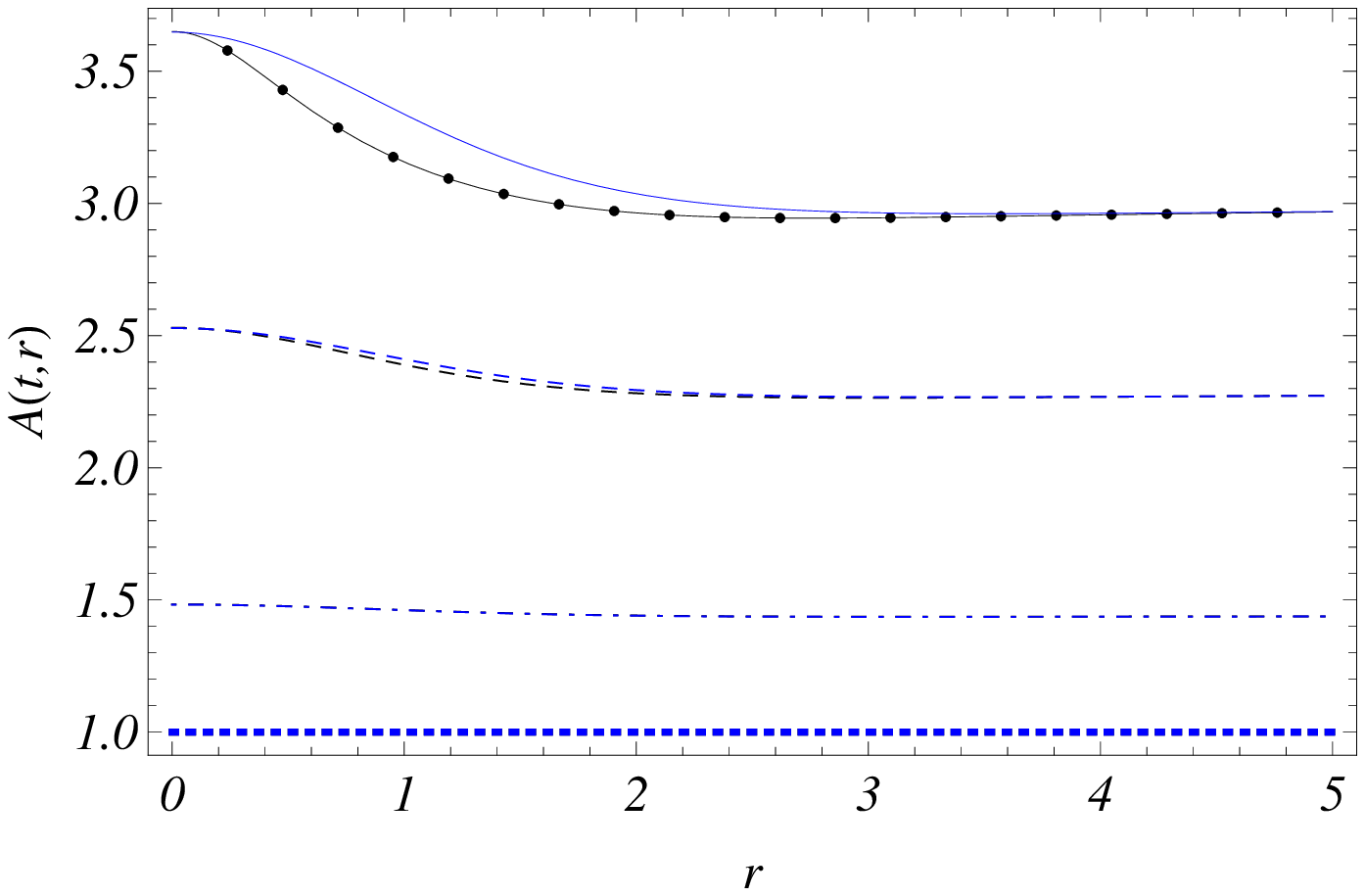,width=8cm}
\epsfig{file=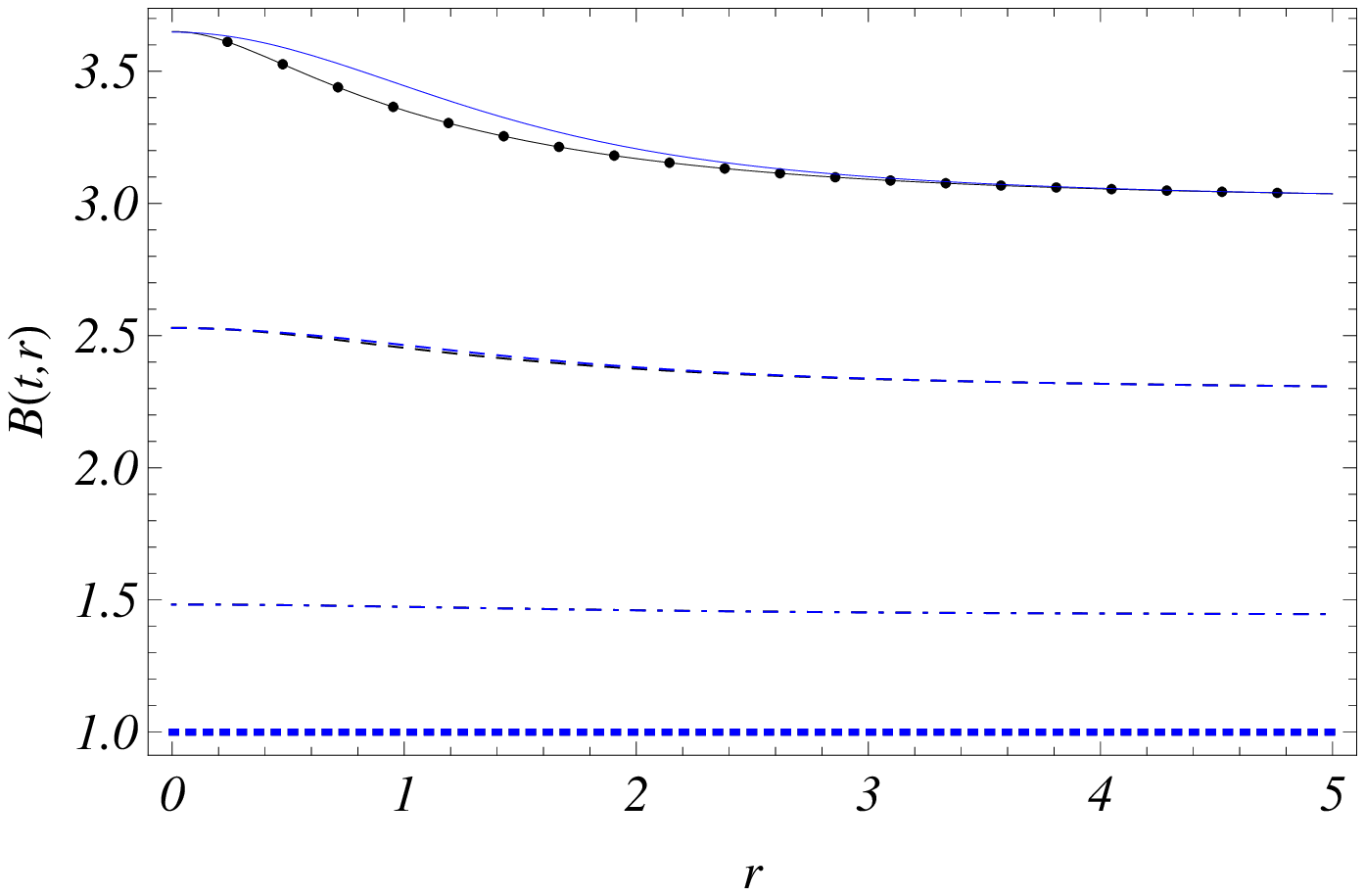,width=8cm}
\caption{Evolution of the scale factors $A(r,t)$ and $B(r,t)$ profile for $\eta=0.1$ (black lines) and $\eta=0.6$ (blue lines) from the time $t_0$ up to the present time $t_p$. A mesh of dots is superimposed to the curve corresponding to $\eta=0.1$, $t=t_p$ to distinguish it from the curve obtained for $\eta=0.6$, $t=t_p$. The profiles shown correspond to $t_0$ (dotted line), $t_p/3$ (dot-dashed line), $2t_p/3$ (dashed line), $t_p$ (solid line). Higher curves correspond to more recent times. Notice that even though the evolution of the scalar field changes significantly as we change $\eta$ from $\eta=0.1$ to $\eta=0.6$ (see Fig.~\ref{fig1}) the corresponding evolution of the scale factors is only slightly changed. The (rescaled) region of accelerated expansion is slightly smaller in the case of smaller $\eta$ as expected due to the more rapidly shrinking core seen in Fig. 1.}\label{fig2}
\end{figure}

In Fig.~\ref{fig2} we show the evolution of the scale factors $A(r,t)$ (upper panel) and $B(r,t)$ (lower panel) profile for $\eta=0.1$  and $\eta=0.6$  from the time $t_0$ up to the present time $t_p$.  Even though the evolution of the scalar field changes significantly as we change $\eta$ from $\eta=0.1$ to $\eta=0.6$ (see Fig.~\ref{fig1}) the corresponding evolution of the scale factors around the present time $t_p$ is mildly modified. This is due to the fact that as we decrease $\eta$, $\lambda$ has to increase more rapidly so that the monopole core contribution to the energy density of the universe at the present time $t_p$ remains unchanged and comparable to the contribution of the matter density. This rapid increase of $\lambda$ leads to a decrease of the monopole core scale compared to the Hubble scale. Thus, even though the expansion rate in the monopole core at the present time is not affected by the value of $\eta$, we cannot decrease $\eta$ by many orders of magnitude below $\eta=0.1$ since this would lead to a monopole core scale that is much smaller than the Hubble scale and therefore would have minor cosmological implications. Thus it is hard to avoid the fine tuning implied by Eqs.~(\ref{tqcond1mat}) and (\ref{tqcond2mat}).

It is now straightforward to obtain the Hubble expansion rate corresponding to each scale factor as a function of redshift and of the comoving distance from the monopole center. For example in the case of the scale factor $A(r,t)$ we have
\be
H_A(r,z)=\frac{{\dot A}(r,t(r,z))}{A(r,t(r,z))}\,, \label{hazdef}
\ee
where the function $t(r,z)$ is obtained by solving numerically the equation $A(r,t)=\frac{1}{1+z}$ with respect to $t$.

\begin{figure}[htbp]
\epsfig{file=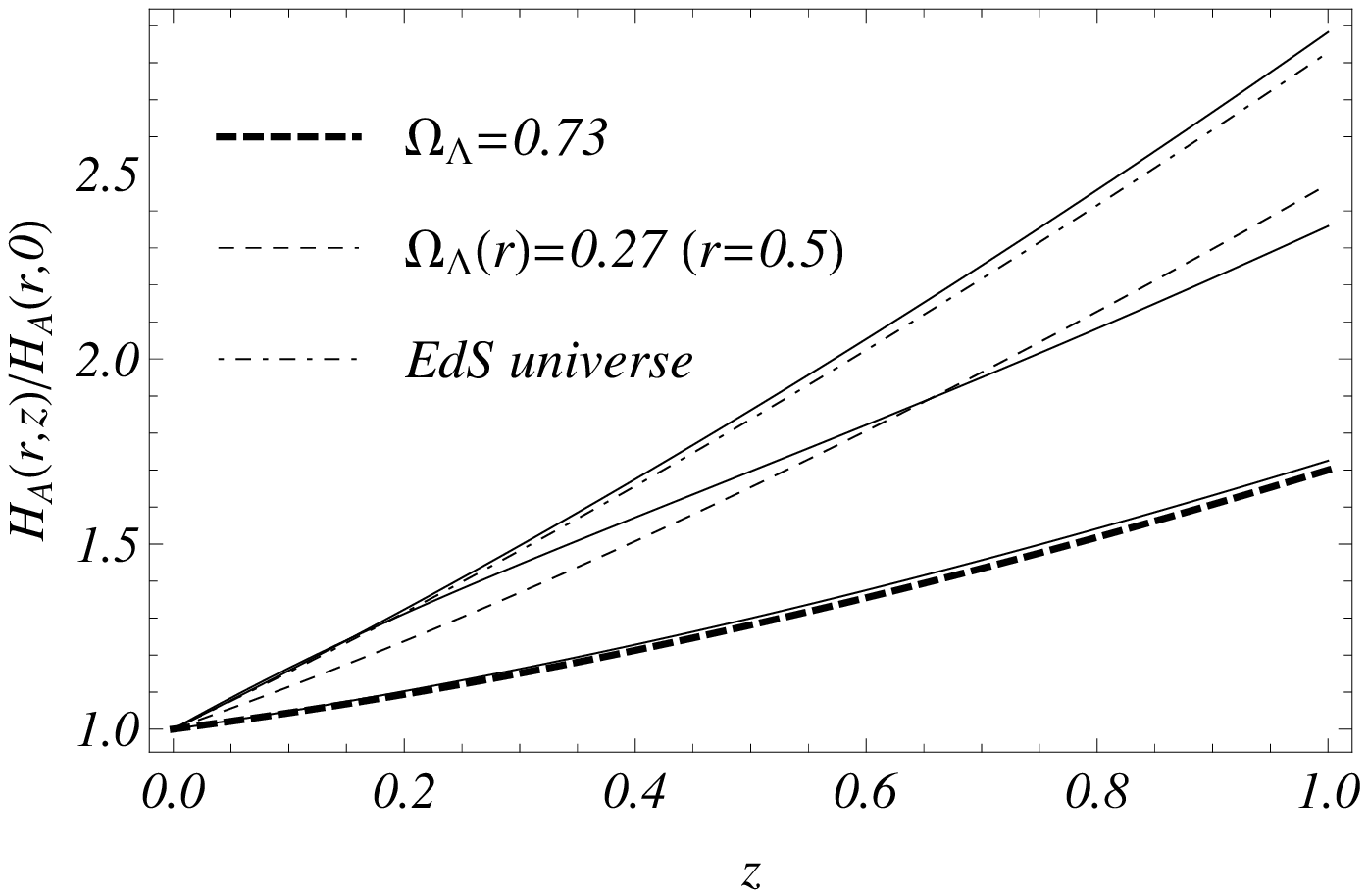,width=8.0cm}
\epsfig{file=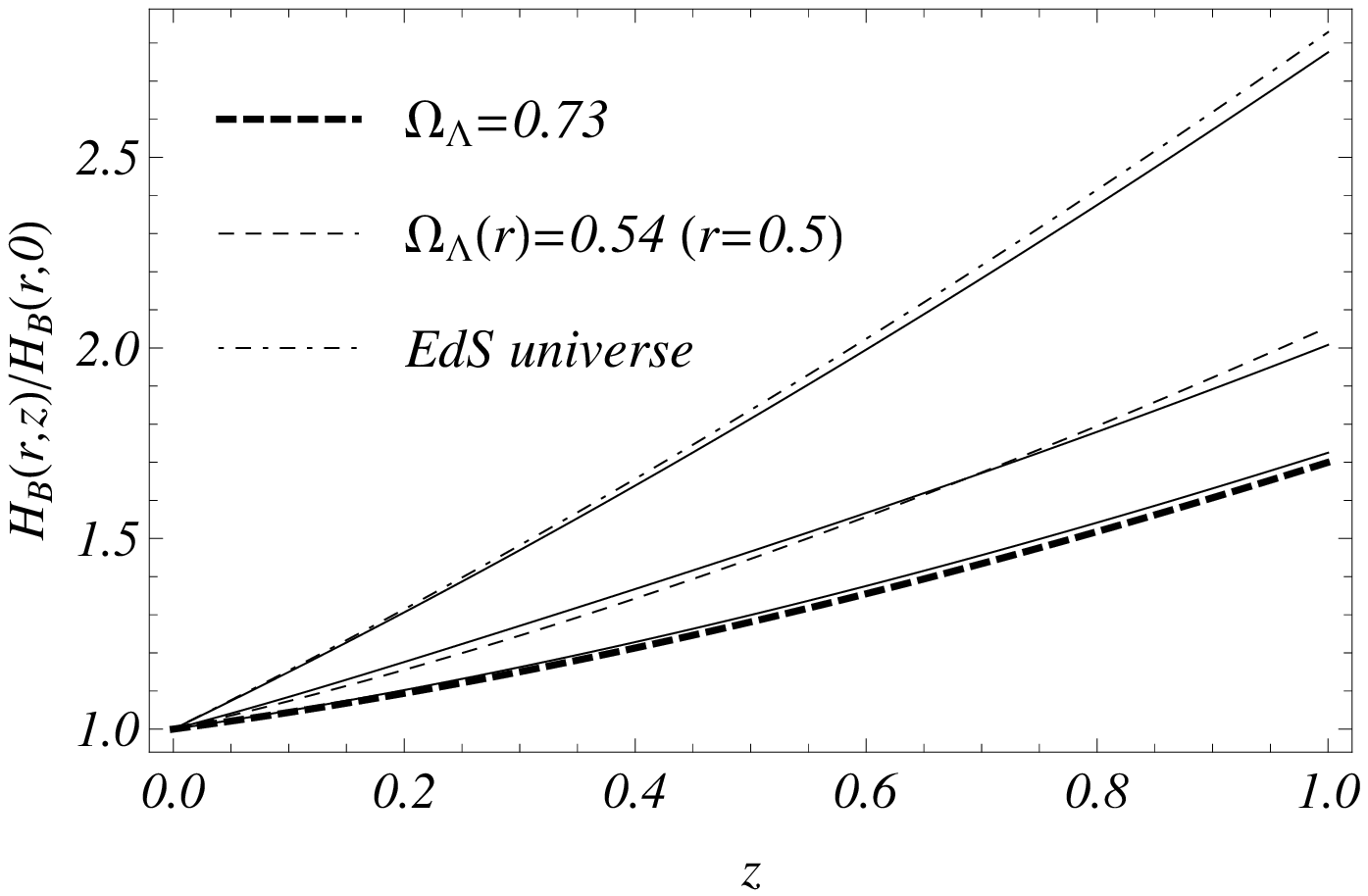,width=8.0cm}\caption{Ratios $H_A(r,z)/H_A(r,0)$ and $H_B(r,z)/H_B(r,0)$ for $r=0$ (lower solid line), $r=0.5$ and $r=5$ (upper solid lines) along with the corresponding $H_\Lambda(r,z)/H_0$ for $\Omega_\Lambda=0.73$ (thick, dashed line) and for $\Omega_\Lambda=0$ (dot-dashed line). Also included for comparison $H_\Lambda(r,z)/H_0$ for the best fit values of $\Omega_\Lambda(r)$ shown in Fig.~\ref{fig4}: $\Omega_\Lambda(0.5)\simeq0.27$ (top panel) and $\Omega_\Lambda(0.5)\simeq0.54$ (bottom panel). We have set $\eta=0.1$.}\label{fig3}
\end{figure}

The derived forms of the Hubble expansion rates $H_A(r,z)$, $H_B(r,z)$ are well fit by the \lcdm form of $H_\Lambda(r,z)^2 = H_0^2 \left(\Omega_\Lambda (r) + (1-\Omega_\Lambda (r))(1+z)^3\right)$ where $\Omega_\Lambda(r)$ interpolates smoothly between $\Omega_\Lambda(0)=0.73$ and $\Omega_\Lambda(r\gg 1)\simeq 0$. This is demonstrated in Fig.~\ref{fig3} where we show $H_A(r,z)/H_A(r,0)$ and $H_B(r,z)/H_B(r,0)$ for $r=0$, $r=0.5$ and $r=5$ along with the corresponding $H_\Lambda(r,z)/H_0$ for $\Omega_\Lambda=0.73$ (thick, dashed line), for $\Omega_\Lambda=0$ (dot-dashed line) and for the best fit $\Omega_\Lambda$ corresponding to $r=0.5$.

\begin{figure}[htbp]
\epsfig{file=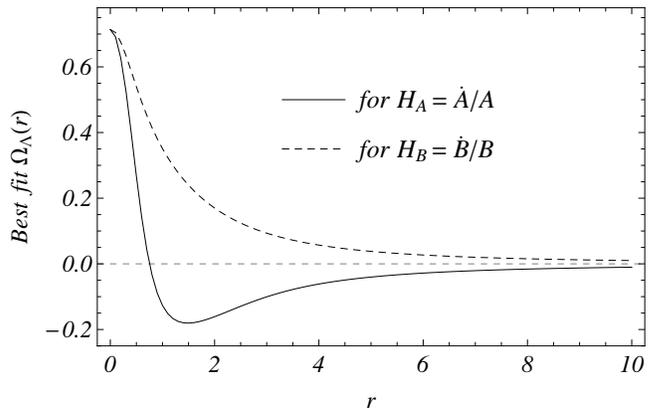,width=8.5cm}\caption{Values of $\Omega_\Lambda(r)$ that provide the best fit to $H_A(r,z)/H_A(r,0)$ (solid line) and to $H_B(r,z)/H_B(r,0)$ (dashed line).}\vspace{1cm} \label{fig4}
\end{figure}

The profiles of $\Omega_\Lambda(r)$ that provide the best fit to $H_A(r,z)/H_A(r,0)$ and to $H_B(r,z)/H_B(r,0)$ are shown in Fig.~\ref{fig4}.
A similar profile for $H_B(r,z)/H_B(r,0)$ has been used in Ref. \cite{Grande:2011hm} to derive detailed observational constraints for inhomogeneous dark energy models. It was argued using qualitative arguments that the expansion rate predicted by topological quintessence could be well described by such a profile. Our numerical analysis confirms this expectation and also derives the precise form of this profile shown in Fig.~\ref{fig4}.

\begin{figure}[htbp]
\epsfig{file=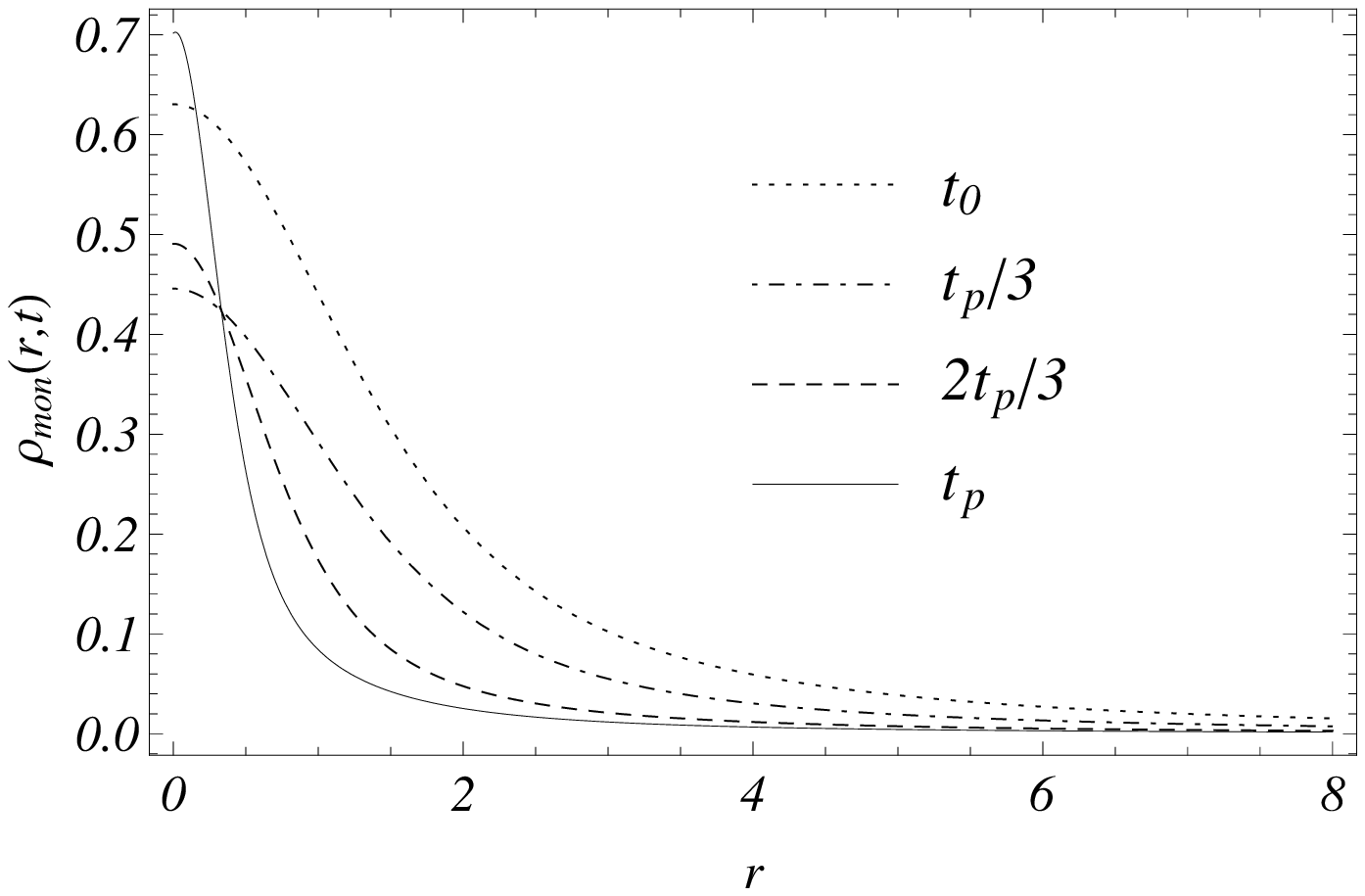,width=8cm}
\epsfig{file=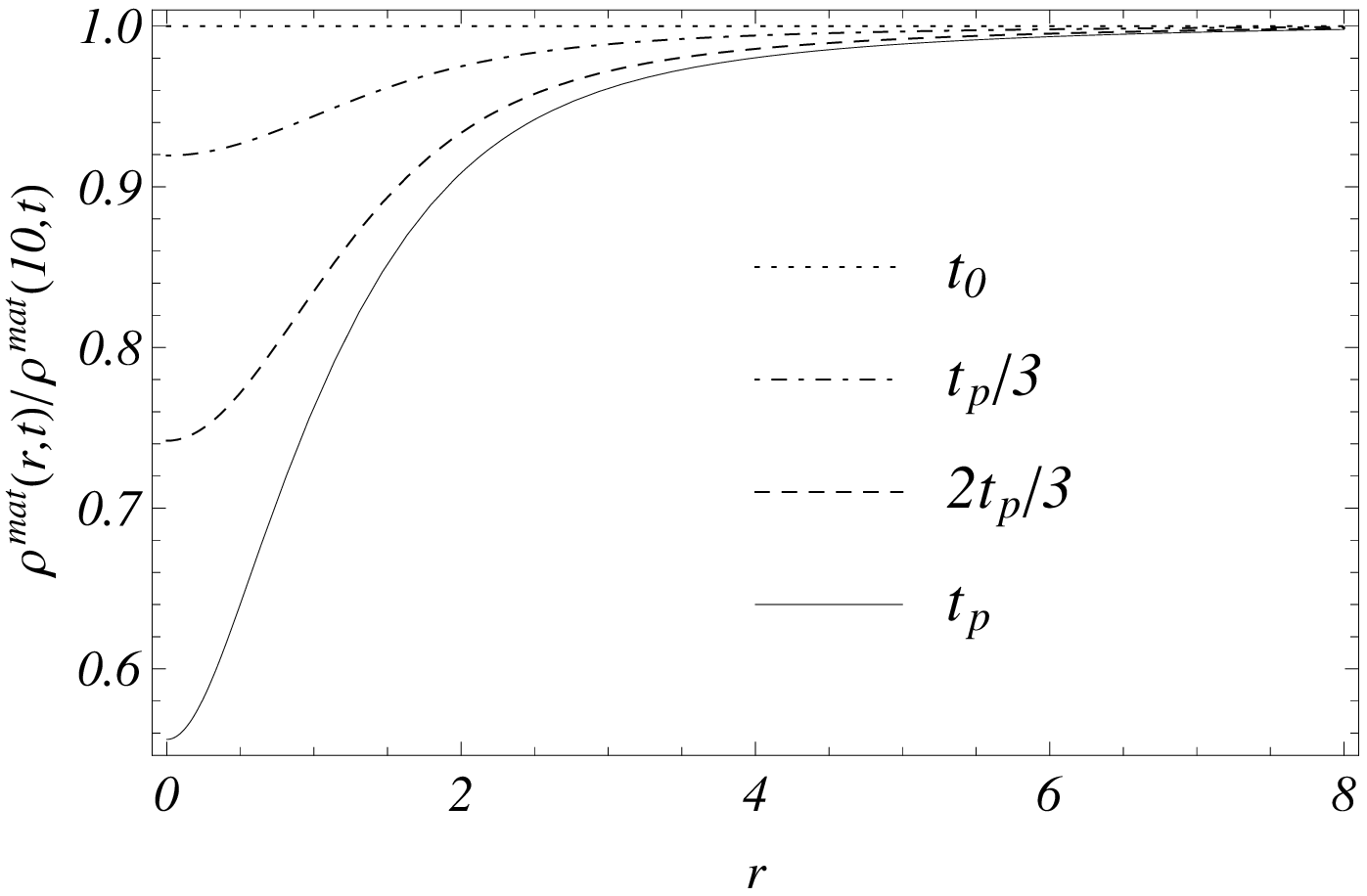,width=8cm}
\caption{Evolution of the dark energy (top panel) and properly normalized matter (bottom panel) density profiles.The profiles shown correspond to $t_0$ (dotted line), $t_p/3$ (dot-dashed line), $2t_p/3$ (dashed line), $t_p$ (solid line). Notice the matter underdensity that develops in the monopole core while the monopole density appears to slowly collapse to the center in comoving coordinates.}\label{fig5}
\end{figure}

The isocurvature nature of the global monopole dark energy inhomogeneity is demonstrated in Fig.~\ref{fig5} where we show the evolution of the dark energy and matter density profiles. The dark energy density profile is defined in Eq.~(\ref{rhomon}). Notice the anticorrelation between dark matter and dark energy inhomogeneities which is expressed by the development of matter underdensity in the monopole core where dark energy is overdense. This nonlinear result confirms the perturbative analysis of Ref. \cite{Sanchez:2010ng} where this anticorrelation was also pointed out. The evolution of the velocity of matter was also evaluated during the simulation and was found to remain very close to $0$ throughout the evolution.

\begin{figure}[htbp]
\epsfig{file=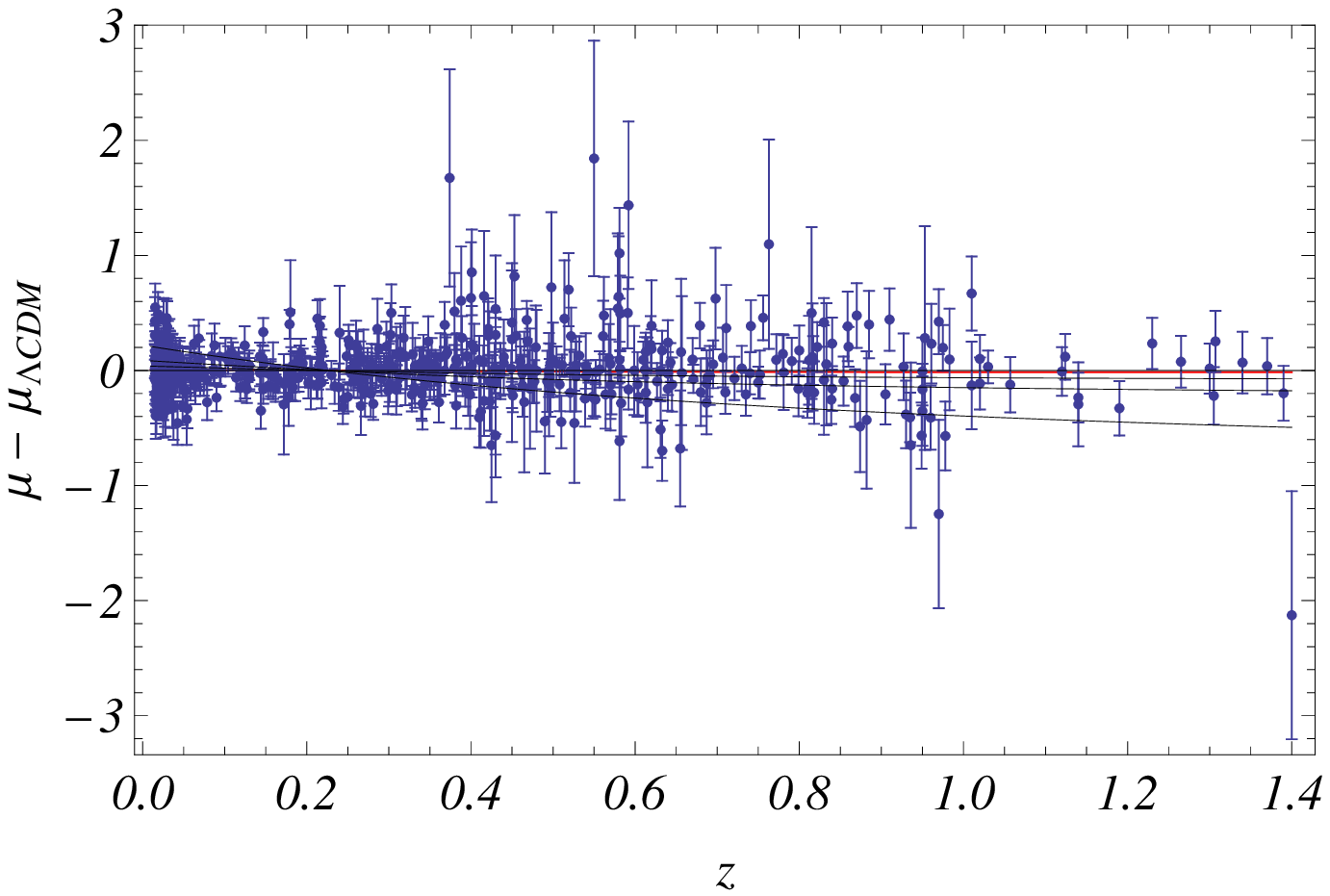,width=8cm}\vspace{0.5cm}
\epsfig{file=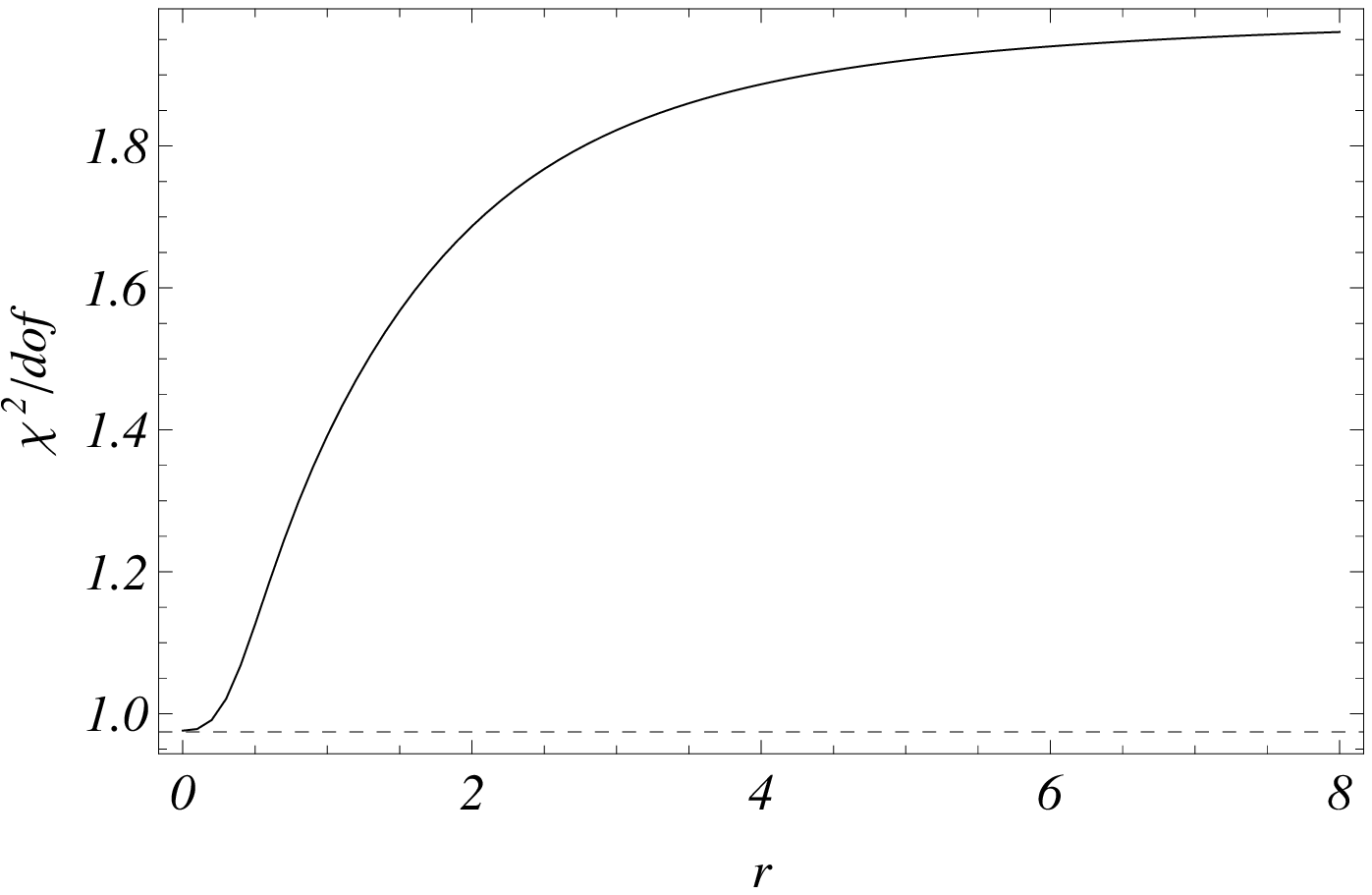,width=8cm}
\caption{{\it Top Panel:} Predicted distance moduli curves (residual with respect to \lcdm best fit) based on the expansion rate for various values of $r$ are shown in the left panel superposed on the Union2 data. The distance moduli curves correspond to $r=0$ (best fit to Union2; red line), $r=0.25$, $r=0.5$, $r=5$ (worst fit corresponding to EdS matter dominated universe away from the monopole core). {\it Bottom Panel:} The quality of fit of the expansion rate $H_B(r,z)$ to the Union2 data expressed through $\chi^2(r)/d.o.f.$. The dashed line corresponds to $\Lambda$CDM in both panels.}\label{fig6}
\end{figure}

Using the derived forms of the Hubble expansion rates it is now straightforward to obtain the luminosity distance using $H_B(r,z)$. This luminosity distance may be fit to the Union2 data \cite{Amanullah:2010vv} and the value of $\chi^2(r)/d.o.f.$ ($\chi^2$ per degree of freedom) may be evaluated as a function of $r$. The predicted distance moduli curves (residual with respect to \lcdm best fit) for various values of $r$ are shown in Fig.~\ref{fig6} (top panel) superposed on the Union2 data. Notice how the agreement deteriorates as $r$ increases away from the core $r\simeq 1$ to the Einstein-de Sitter (EdS) matter dominated region. This is quantitatively expressed in Fig.~\ref{fig6} (bottom panel) where we show the $\chi^2(r)/d.o.f.$ which changes from a value less than 1 close to the center to a value close to 2 away from the core. Thus a global monopole with a Hubble scale core provides a good fit to the Union2 data. A corresponding analysis of spherical inhomogeneous dark energy, based on a toy model but taking into account the detailed inhomogeneous structure of the dark energy profile may be found in Ref.~\cite{Grande:2011hm}.

\section{Conclusion}\label{concl}
We have introduced a simple new mechanism called {\it topological quintessence} that constitutes a generic generalization of $\Lambda$CDM. Instead of breaking time translation invariance of the cosmological constant which occurs in most studied generalizations of \lcdm (eg. the usual quintessence), topological quintessence involves breaking {\it space translation invariance} of the cosmological constant. This breaking is achieved via a recent low energy phase transition which gives rise to topological defects with Hubble scale core and energy density in the core comparable to the present matter density.

We have studied in some detail the case of spherical global defects (global monopoles) while our mechanism can be easily generalized to different symmetries and different field varieties. We have shown that for proper values of field theory parameters shown in Eqs.~(\ref{tqcond1mat}) and (\ref{tqcond2mat}) topological quintessence is consistent with the Union2 Type Ia supernovae data and predicts recent accelerating expansion on Hubble scales very similar to the one predicted by $\Lambda$CDM. Due to the cosmologically large scale of the dark energy inhomogeneity there is also no breaking of the Copernican principle.

The recent nature of the phase transition involved in topological quintessence allows for a matter era that is for the most part indistinguishable from that of the standard \lcdm model. Therefore only the large scale low multipole part of the CMB spectrum is expected to be modified. A general study of such anticipated observational consequences in this class of models may be found in Ref. \cite{Grande:2011hm}. In order to implement the general observational constraints obtained in Ref. \cite{Grande:2011hm} to the particular case of topological quintessence, we need to establish the correspondence between the general parameters of Ref. \cite{Grande:2011hm} and the parameters of the topological quintessence models. This correspondence may be summarized as follows:
\begin{itemize}
\item
The size of the dark energy inhomogeneity $r_0$ of Ref. \cite{Grande:2011hm} should ne identified with the scale of the global monopole core $\delta \simeq \lambda^{-1/2} \eta^{-1}$.
\item
The normalized dark energy density in the center of the inhomogeneity ($\Omega_{Xin}$ of Ref. \cite{Grande:2011hm}) should be identified with the normalized potential energy of the global monopole center  $\frac{\rho^{core}}{\rho_c} \simeq \frac{\lambda \eta^4}{4\; \rho_c}$ where $\rho_c$ is the present critical density for flatness.
\item
The distance $r_{obs}$ of the observer from the center has an identical meaning in both approaches.
\end{itemize}

The constraints on the parameters $r_0$, $\Omega_{Xin}$ and $r_{obs}$ obtained in Ref. \cite{Grande:2011hm} may be summarized as follows:
The Union2 SnIa data constrain $r_0$, $\Omega_{Xin}\equiv 1-\omm$ to $15Gpc> r_0>1.76 Gpc$ (the upper bound corresponds to homogeneous \lcdm) $0.8>\Omega_{Xin}>0.54$ at the $3\sigma$ level (Fig. 2 of Ref. \cite{Grande:2011hm}). The $3\sigma$ range of $r_0$ corresponds to a range of the parameter $\alpha$ (defined above eq. (\ref{condmat1})) as $5>\alpha>0.6$.
Using the $3\sigma$ range of the parameters $\Omega_{Xin}$ and $r_{obs}$  (or equivalently $\Omega_{0m}$ and $\alpha$ defined above eq. (\ref{condmat1})) and equations (\ref{tqcond1mat}) and (\ref{tqcond2mat}) we obtain the observationally determined $3\sigma$ range of the model parameter $\lambda$ and $\bar \eta$. We find at the $3\sigma$ level \be 4\times 10^{-117} h^2>\lambda >1.1\times 10^{-122} h^2 \ee and \be 0.63>{\bar \eta}>0.02 \ee.

The value of $r_{obs}$ is very weakly constrained by the SnIa data and it can be close to the $r_0$ (Fig. 6 of Ref. \cite{Grande:2011hm}) especially for large inhomogeneity scales $r_0$ ($4Gpc$ or larger).  The value of $r_{obs}$ is more severely constrained by the smallness of the observed CMB dipole. For $r_0$ up to $7 Gpc$, $r_{obs}$ is constrained to be less than $110Mpc$ (Fig. 7 of Ref. \cite{Grande:2011hm}) while the CMB dipole constraint on $r_{obs}$ is alleviated if the inhomogeneity reaches the distance to the last scattering surface ($r_0 \simeq 13 Gpc$). An extended study of these constraints as well as derivation of specific observational signatures of topological quintessence is an important next step but is beyond the scope of the present study.

The following directions could lead to useful extensions of the present project:
\begin{itemize}
\item
A more detailed investigation of the above mentioned observational consequences and their relation to CMB anomalies \cite{Copi:2010na} and other puzzles \cite{Perivolaropoulos:2008ud} of the standard \lcdm model \cite{lcdmrev}.
\item
The consideration of multi-defect configurations could reveal interesting new effects and observational signatures. Such signatures are expected to include Hubble scale velocity flows experienced by off center observers, aligned low multipole CMB moments in multi-defect configurations, directional variations of fundamental constants due to possible coupling to the defect scalar field etc.
\item
The application of our analysis to the case of non-minimally coupled scalar field ({\it extended topological quintessence}). In this case we anticipate a more drastic reaction of the matter profile at the monopole core.
\item
The consideration of alternative topological defects (a global string, a wall or a gauged defect). The identification of distinct observational signatures for each defect geometry constitutes an useful probe of this class of models.
\item
Finally our numerical tools may also be used with proper initial conditions to investigate the effect of dark energy (minimally and non-minimally coupled) on the profiles of bound matter structures like clusters of galaxies. This analysis is currently in progress.
\end{itemize}

\section*{Numerical Analysis Files}
The Mathematica files used to produce the figures of this study may be downloaded from http://leandros.physics.uoi.gr/topquint .

\section*{Appendix}
In the presence of matter and far away from the monopole core ($r\gg \delta$), using Eq.~(\ref{dyneq0}) we find that the approximate form of the scale factors $A(r,t)$ and $B(r,t)$ satisfy the equations:
\begin{widetext}
\ba
&&\frac{3}{2} \beta^{3/2} \left( \sqrt{\beta^{-1}\frac{B(r,t)}{B_{mat}}}\sqrt{1+\beta^{-1}\frac{B(r,t)}{B_{mat}}}-\sinh^{-1}\sqrt{\beta^{-1}\frac{B(r,t)}{B_{mat}}}\right) \simeq  \frac{t}{t_{mat}}
\\
&&A(r,t)\simeq  B(r,t)
\label{scfacfarmon}
\ea
\end{widetext}
where  $\beta \equiv \frac{B_{eq}}{B_{mat}}$, $B_{eq}(r,t_{eq})$ is the scale factor at the time when the density of matter is equal to the energy density of the monopole ($\rho^{mat}(t_{eq})=\rho^{mon}(t_{eq})$, $B(r,t_{eq})\equiv B_{eq}$) and $B_{mat}$ is the scale factor at an early time $t_{mat}$ when matter dominates ($t_{mat} \ll t_{eq}$).  It is easy to check that the above expression reduces to the anticipated forms in the matter dominated ($B \ll B_{eq}$) and in the monopole dominated ($B > B_{eq}$) eras. These expressions are of the form
\ba
A(r,t)&\simeq & B(r,t) \simeq \left(\frac{t}{t_{mat}}\right)^{2/3} \;\;,\;\;  B\ll B_{eq} \\
A(r,t)&\simeq & B(r,t) \simeq \sqrt{\frac{8\pi \eta^2}{3\; m_{Pl}^2}}\frac{t}{r} \;\;,\;\; B\gg B_{eq}
\label{bmonera}
\ea
where the rescaling of the scale factor in the monopole era $\frac{1}{\sqrt{3}}$ compared to Eq.~(\ref{scfmonera2}) is due to the matching with the preceding matter era.

\end{document}